\documentclass[12pt]{iopart} %\documentclass{article} 
\usepackage{epsfig,amssymb,mathrsfs} %\documentclass{report}

\def\beq{\begin{equation}} 
\def\eeq{\end{equation}} \def\bea{\begin{eqnarray}} 
\def\eea{\end{eqnarray}} \def\d{{\mathrm{d}}} 
  
 \newfont{\cursive}{pzcmi at 
9pt} \def\~t{\tilde{t}}

\begin{document}

\title[Causal and quasi-local horizons and horizon-entropy under conformal rescalings]{The horizon-entropy increase law for causal and quasi-local horizons and conformal field redefinitions} 
\author{Alex B. Nielsen \\ Max-Planck-Institut f\"ur  
Gravitationsphysik, \\ 
Albert-Einstein-Institut, \\ 
Am M\"uhlenberg 1, D-14476 Golm, \\ Germany} 
\author{Valerio Faraoni \\ Physics 
Department, Bishop's University \\ 2600 College Street \\ 
Sherbrooke, Qu\'ebec, Canada J1M~1Z7}

%\begin{document} \maketitle 
\begin{abstract} 
We explicitly prove 
the horizon-entropy increase law for both causal and quasi-locally defined 
horizons in scalar-tensor and $f(R)$ gravity theories. Contrary 
to causal event horizons, future outer trapping horizons are not 
conformally invariant and we provide a modification of trapping 
horizons to complete the proof, using the idea of generalised 
entropy. This modification means they are no longer foliated by 
marginally outer trapped surfaces but fixes the location of the 
horizon under a conformal transformation. We also discuss the 
behaviour of horizons in ``veiled'' general relativity and show, 
using this new definition, how to locate cosmological horizons in 
flat Minkowski space with varying units, which is physically 
identified with a spatially flat FLRW spacetime. 
\end{abstract}


%\tableofcontents

%********************************************************************** 
\section{Introduction} 
%**********************************************************************

The entropy of a black hole is not always given simply by 
one quarter of its area. In alternative theories of 
gravity, such as Brans-Dicke or $f(R)$ theories, the horizon-entropy of 
the black hole is given by a more complicated function of the 
black hole geometry and possible horizon fields. In such cases, 
ensuring that the entropy of the black hole is non-decreasing is 
not equivalent to ensuring that the area is non-decreasing. A 
number of authors have been able to prove an equivalent of Hawking's 
area increase theorem for black hole event horizons in several alternative theories \cite{Jacobson:1995uq, Kang:1996rj, 
Ford:2000xg}.

Quasi-local horizons also have an area increase law 
\cite{Hayward:1993wb}. The thermodynamic properties of apparent 
horizons and their quasi-local associates, dynamical and trapping 
horizons, have been investigated in \cite{Hayward:1993wb, 
Ashtekar:2004cn}, and \cite{Nielsen:2008cr}. In Einstein gravity 
the area of a trapping horizon is guaranteed to  be non-decreasing
 if the 
null energy condition is satisfied. This result for the area is true even 
in alternative gravity situations, provided the null convergence condition is substituted for the null energy condition. But this does not 
guarantee that the horizon-entropy of the trapping horizon is non-decreasing.

In this article we will examine situations where the 
horizon-entropy is not one quarter of the area. We examine both causal horizons and quasi-local horizons. Causal horizons are the null causal boundaries of a given spacetime region and include event horizons, which are the past causal boundary of future null infinity. Quasi-local horizons include dynamical and trapping horizons, but we will also investigate a new 
definition, closely related to that of a trapping horizon, that 
satisfies a horizon-entropy increase law in a range of situations \cite{Nielsen:2010gm}. This 
new surface  has the important property that it reduces to 
that 
of a trapping horizon in cases where the horizon-entropy is one 
quarter of the area. It therefore retains all of the previous results 
for trapping horizons in the case of Einstein gravity and extends 
their validity to other theories.  We extend the results in 
\cite{Nielsen:2010gm} to a much wider class of gravity theories, 
including scalar tensor theories and $f(R)$ theories and also 
extend the results to a much wider class of horizons, including 
ones that are not necessarily spherically symmetric. In addition, we
derive a corresponding horizon-entropy equation for causal 
horizons that unifies many of the previous results that have 
appeared in the literature.  

This new horizon  definition has the property that under a 
conformal 
transformation of the metric, its location and in particular its 
relation to the event horizon is unchanged. This is not true of 
trapping horizons. The use of conformal transformations is fairly 
common in the study of gravity theories. This is particularly 
true in string theory where conformal transformations are used to 
relate the string frame, with a non-minimally coupled dilaton 
field, to the Einstein frame.\footnote{Several authors have 
already noted that they should more properly be called 
``representations''  rather than ``frames'' 
\cite{Ford:2000xg,Deruelle:2010ht}.} It has been argued in the 
literature that, classically, the two frames are physically 
equivalent \cite{Dicke, Flanagan:2004bz, Faraoni:2006fx, 
Deruelle:2010ht}. This physical equivalence suggests that the new 
horizons should be preferred to trapping horizons if these 
surfaces are to have physical significance, such as a role in 
black hole thermodynamics and Hawking radiation.

The conformal transformation rescales lengths and areas as 
measured by the metric. The physical effect of this rescaling is, 
for example, to change the meaning of mass since the norm of the 
four-momentum $p^{a}p_{a}$ will no longer be constant from point 
to point or from time to time. The importance of running units in 
making the correspondence was emphasised in \cite{Dicke, 
Faraoni:2006fx}. The example of Einstein gravity in a frame 
where gravity is not minimally coupled to the matter fields was 
explicitly examined in \cite{Deruelle:2010ht}. In this case, 
where there are no ``fundamental'' scalar fields, the 
observational predictions are still exactly the same in two 
different frames. The authors of Ref.~\cite{Deruelle:2010ht} use 
the term ``veiled general relativity'' to describe this 
situation.

The plan of this paper is as follows: 
Sec.~\ref{sec:generalentropy} provides background material on 
horizon-entropy in modified theories of gravity. 
Sec.~\ref{sec:secondlaw} examines the various proofs for the 
increase of this horizon-entropy for both causal horizons 
and quasi-local horizons such as trapping horizons. The proofs are 
discussed for Einstein gravity, Brans-Dicke gravity, and general 
scalar-tensor and $f(R)$ gravity theories.  Here we see that 
trapping horizons, as commonly defined, can only guarantee 
increase of horizon-entropy in the case of Einstein gravity. However, for the modification given in 
eqs.~(\ref{modtraphorconds}), the horizon-entropy law can be 
guaranteed in a large class of other theories. This modification makes the location of the 
geometrically defined horizon invariant under a conformal 
transformation, as we discuss in 
Sec.~\ref{sec:conformaltransformations}. This allows us to locate 
invariantly defined horizons in conformally equivalent spacetimes 
and we demonstrate this for cosmological horizons in 
Sec.~\ref{sec:cosmologicalhorizons}. Sec.~\ref{sec:conclusion} 
contains a discussion and the conclusions.

%*************************************************************************** 
\section{Horizon-entropy for general gravity theories} 
%*************************************************************************** 
\label{sec:generalentropy}

There are several ways to derive the entropy that should be 
associated with a black hole horizon. For a static spacetime one 
can make use of the Euclideanised action. This was used in 
\cite{Callan:1988hs} to show that static black holes do not obey 
$S=A/4$ to linear order in a particular model of second order 
curvature corrections derived from string theory. This technique 
was later generalised to all orders for Lagrangians that are an 
arbitrary function of the Riemann tensor by Visser 
\cite{Visser:1993qa, Visser:1993nu} who derived the formula  
\beq 
S \label{visserentropy} = \frac{A_{H}}{4} + 
4\pi\int_{H}\frac{\partial{\mathscr{L}}_{m}}{\partial R_{abcd}} \, 
g_{ac}^\perp g_{bd}^\perp\, \sqrt{q} \, \d^{2}x \,, 
\eeq 
where 
integrations should be taken over closed two-spheres, $H$, with 
metric $q_{ab}$, while $g_{ab}^\perp$ is the symmetric metric of 
the two-dimensional subspace orthogonal to these surfaces, spanned by null vectors $l^{a}$ and $n^{a}$ such that ${g_{ab}^\perp} = -l_{a}n_{b} - n_{a}l_{b}$ with $n^{a}l_{a}=-1$.  ${\mathscr{L}}_{m}$ is 
the ``matter'' Lagrangian density, which can be constructed as 
the total Lagrangian density minus the Einstein-Hilbert term.

Alternatively, one can require the validity of the first 
law for Killing horizons of any diffeomorphism-invariant theory. 
This was done in \cite{Wald:1993nt} and gives the result 
\beq \label{waldentropy} 
S = -2\pi\int_{H}\frac{\partial 
{\mathscr{L}}}{\partial R_{abcd}} \, 
\hat{\varepsilon}_{ab}\hat{\varepsilon}_{cd} \, \sqrt{q} \, 
\d^{2}x + \mathrm{higher\hspace{0.1cm} derivative\hspace{0.1cm} 
terms}, 
\eeq 
where $\hat{\varepsilon}_{ab}$ is the antisymmetric 
binormal form for the surface $H$, 
$\hat{\varepsilon}_{ab}=l_{a}n_{b} - n_{a}l_{b}$ and $ {\mathscr{L}} 
$ is the full Lagrangian density. The higher derivative terms 
arise for theories that depend on derivatives of the Riemann 
tensor and we will ignore them here. The equivalence of this 
formula with (\ref{visserentropy}) is obtained by the relation 
$\hat{\varepsilon}_{ab}\hat{\varepsilon}_{cd} = g_{ad}^\perp \, 
g_{bc}^\perp - g_{ac}^\perp \, g_{bd}^\perp$.

What is needed for these formulae is a choice of spacelike 
surface $H$, knowledge of how the Lagrangian density ${\mathscr{L}}$ 
depends on the Riemann tensor, and knowledge of the local 
geometry and fields at the surface $H$.  The horizon-entropy has 
the form of an integral over the two-dimensional surface of a two-form, $S=\int_{H} s_{ab}$ with \beq s_{ab} = 
-2\pi\frac{\partial {\mathscr{L}}}{\partial R_{cdef}} \, 
\hat{\varepsilon}_{cd}\hat{\varepsilon}_{ef}\varepsilon_{ab} \,, 
\eeq which is just a scalar quantity times the area two-form 
$\varepsilon_{ab}$ of the surface $H$. In principle a horizon-entropy two-form can be associated with each point of the horizon, although it depends on which two-surface it is associated with. For the normal 
Einstein-Hilbert action of Einstein gravity, where the ``matter'' 
Lagrangian is zero and hence the Visser horizon-entropy is trivial, we 
have \beq {\mathscr{L}} = \frac{R}{16\pi} \,, \eeq \beq 
\frac{\partial {\mathscr{L}}}{\partial R_{abcd}} = \frac{1}{16\pi} \, 
\frac{1}{2}\left( g^{ac} g^{bd} - g^{ad} g^{bc} \right) \,, 
\eeq 
thus, 
\beq 
s_{ab} = -2\pi \left(\frac{1}{16\pi}\right) 
\hat{\varepsilon}^{cd} \hat{\varepsilon}_{cd} \, \varepsilon_{ab} 
\,, 
\eeq 
and therefore, since $\hat{\varepsilon}^{cd} 
\hat{\varepsilon}_{cd} = -2$, 
\beq 
S = \frac{A}{4} \,, 
\eeq 
$A$  being the area of $H$. In the case of scalar-tensor gravity 
\cite{BransDicke}, we have 
\beq 
{\mathscr{L}} = 
F(\phi)\frac{R}{16\pi} + 
\mathrm{other\hspace{0.1cm}terms\hspace{0.1cm}independent\hspace{0.1cm}of\hspace{0.1cm}Riemann} 
\eeq 
and thus 
\beq 
s_{ab} = \frac{F(\phi)}{4} \, \varepsilon_{ab} \,. 
\eeq 
When $F(\phi)$ is constant over the horizon, for 
example for a spherically symmetric surface, we have 
\beq 
S = 
\frac{F(\phi)A}{4} 
\eeq 
(cf. \cite{STentropy}), while in 
the case of $f(R)$ gravity \cite{f(R)reviews}, we have 
\beq 
{\mathscr{L}} = \frac{f(R)}{16\pi} 
\eeq and thus 
\beq 
s_{ab} = 
\frac{f'(R)}{4} \, \varepsilon_{ab} \,. 
\eeq 
Again, in the case 
where $f'(R)$ is constant over $ H $, this gives \beq S = 
\frac{f'(R)A}{4} \,. \eeq

In all these cases the horizon-entropy has the form $S=WA$ for some scalar function $W$. The horizon-entropy in \cite{Wald:1993nt} was explicitly derived to apply to 
Killing horizons in a stationary spacetime. It was suggested in 
\cite{Iyer:1994ys} that in certain cases the entropy could also 
take this form for non-stationary situations. We will henceforth 
refer to~(\ref{waldentropy}) as the horizon-entropy, without 
prejudice to the question of whether it represents a true entropy or not in dynamical situations. The question then 
arises as to what kind of surface this horizon-entropy can be applied to. In 
non-stationary situations the event horizon does not in general 
coincide with the trapping horizon even though both satisfy an 
area increase law in Einstein gravity. In the 
next section we will consider to what extent the horizon-entropy 
satisfies an increase law for non-stationary surfaces.

%********************************************************************** 
\section{The second law of black hole mechanics} 
%********************************************************************** 
\label{sec:secondlaw}

Let us consider a three-dimensional surface that can be foliated 
by closed spacelike two-dimensional surfaces (such an object 
could be an event horizon, a trapping horizon, or even something 
else). In a four-dimensional Lorentzian signature spacetime the 
spacelike two-surfaces have null normals $l^{a}$ and $n^{a}$ that 
are unique up to scalings. $l^{a}$ and $n^{a}$ are conventionally referred to as the out-going future-directed null normal and in-going future-directed null normal respectively. The tangent $r^{a}$ to the 
surface, which is normal to the spacelike two-surfaces, can be 
written everywhere as a linear combination of $l^{a}$ and 
$n^{a}$, 
 
\beq \label{rwithBandC} r^{a} = Bl^{a} + Cn^{a} \, .
\eeq 
For a Killing horizon, or a non-stationary event horizon, or 
general causal horizon, 
$r^{a}$ would be the generators of the horizon, and would 
therefore be null with either $B=0$ or $C=0$ and in the case of a 
dynamical horizon we would have $B>0$ and $C<0$. A future outer trapping horizon can have any sign for $C$. The signature of 
the three-dimensional surface is just given by the norm 
squared of $r^{a}$, 
\beq \label{signature} 
r^{a} r_{a} = 2BCl^{a}n_{a}  
\eeq 
where, for future directed null normals, $l^{a}n_{a}$ is negative.  The discussion here will follow that of 
\cite{Hayward:1993wb} where $l^{a}$ and $n^{a}$ can be chosen 
such that $B$ and $C$ above are constant on the two-dimensional surfaces.   To fix a direction on 
the three-dimensional horizon surface we can 
choose  $B>0$. The horizon will then be spacelike if $C<0$, null  
 if $C=0$, and timelike if $C>0$. The horizon-entropy will in all 
these cases be  non-decreasing  if 
\beq \int {\cal{L}}_{r}s_{ab} \geq 0 \,, 
\eeq 
with 
${\cal{L}}_{r}$ the Lie derivative along $r^{a}$. Now one 
can look at how the horizon-entropy two-form $s_{ab}$ varies as 
one moves along integral curves of $r^{a}$ from one spacelike 
two-surface to another. 
\beq 
{\cal{L}}_{r}s_{ab} =  B{\cal{L}}_{l}s_{ab} + 
C{\cal{L}}_{n}s_{ab} \,. 
\eeq 
Since the 
entropy two-form can be written as $s_{ab}=W \, 
\varepsilon_{ab}$, this equation is equivalent to 
\beq 
{\cal{L}}_{r}s_{ab} = \left[B\left({\cal{L}}_{l}W + 
W\theta_{l}\right)+C\left({\cal{L}}_{n}W + 
W\theta_{n}\right)\right]\varepsilon_{ab} \,, 
\eeq 
where we define  the expansion $\theta_{l}$ by  
${\cal{L}}_{l}\varepsilon_{ab} = 
\theta_{l}\varepsilon_{ab}$. Determining, or defining, that the 
signs of the scalar terms in ${\cal{L}}_{r}s_{ab}$ combine to 
give an overall non-negative result implies that the 
entropy two-form is non-decreasing in the direction of $r^{a}$ 
everywhere on $H$ and thus the horizon-entropy is non-decreasing 
along the three-dimensional surface in question. For causal 
horizons, this just reduces to the requirement that $ 
\varepsilon^{ab}{\cal{L}}_{l}s_{ab}$ be non-negative, which can 
be related to the equations of motion and an energy condition. In 
situations where the horizon is not null, as we will see below, 
the sign of the $ C\varepsilon^{ab}{\cal{L}}_{n}s_{ab}$ term can 
also be evaluated in a similar manner.

\subsection{Einstein gravity} 
%**********************************************************************

In the usual case of Einstein gravity we have $s_{ab} = 
\varepsilon_{ab}/4$ and $S = A/4$. In this case the horizon-entropy 
increase law for event horizons is just the area theorem of 
Hawking \cite{Hawking:1971tu, Hawking:1973uf}. Since it does not 
affect the sign of the change in entropy, henceforth we will 
incorporate the factor of $4$ into $A$ for notational 
convenience.

For the case of quasi-local horizons, foliated by marginally 
outer trapped surfaces with outgoing null expansion  
$\theta_{l} = 0$ and ingoing null expansion negative, 
$\theta_{n} < 0$, the variation of the horizonentropy two-form is simply 
\bea 
\label{grtraphorvarent} {\cal{L}}_{r}\varepsilon_{ab} & 
= & B{\cal{L}}_{l}\varepsilon_{ab} + 
C{\cal{L}}_{n}\varepsilon_{ab} \nonumber \\ & = & C \, 
\theta_{n} \varepsilon_{ab}\,. 
\eea 
If $C$ is  assumed to be negative, the area-entropy 
is non-decreasing 
without 
further assumptions. This is the case considered for dynamical 
horizons in \cite{Ashtekar:2004cn} since dynamical horizons are 
required to be spacelike and by eq.~(\ref{signature}) this 
guarantees $C<0$.

In the more general case of a future outer trapping horizon, 
which can have any signature, the sign of $C$ can be related to 
the energy conditions via the condition that $\theta_{l}$ should 
be zero everywhere on the trapping horizon. The conditions for a 
future outer trapping horizon are \cite{Hayward:1993wb} 
\bea \label{fothconds} 
& \theta_{l} = 0 \,,& \nonumber \\
& \theta_{n} < 0 \,,& \nonumber \\ 
& {\cal{L}}_{n}\theta_{l} < 0 \,.&
\eea 
For a past inner trapping horizon one would interchange the 
$n$'s and the $l$'s and reverse the sign of the inequalities. The 
third condition distinguishes trapping horizons from dynamical 
horizons. The constancy of the expansion $\theta_{l}$ on the 
horizon gives the condition \beq {\cal{L}}_{r}\theta_{l} = 
B{\cal{L}}_{l}\theta_{l} + C{\cal{L}}_{n}\theta_{l} = 0 
\,.
\eeq  
The Raychaudhuri equation for null geodesic congruences 
is 
\beq \label{raychaudhuri} 
{\cal{L}}_{l}\theta_{l} = \kappa_{l} 
\theta_{l} - \frac{1}{2} \, \theta_{l}^{2} - \sigma_{l}^{2} + 
\omega_{l}^{2} - R_{ab}l^{a}l^{b} \,, 
\eeq 
where $\kappa_{l}$ is 
a measure of the failure of $l^{a}$ to be affinely 
parameterised (a ``surface gravity'' \cite{Nielsen:2007ac}), 
$\sigma_l$ is the shear, and $\omega_{l}$ is the vorticity. If 
the null vectors used to define the horizon are derived from a 
double-null foliation (this construction is used in 
\cite{Hayward:1993wb}), then the vorticity vanishes identically 
and for $\theta_{l}=0$ we have \beq \label{40} 
{\cal{L}}_{l}\theta_{l} = - \sigma_{l}^{2} - R_{ab}l^{a}l^{b} 
\,, \eeq and we obtain  
\beq \label{valueC} 
C = 
\frac{B}{{\cal{L}}_{n}\theta_{l}} \left( \sigma_{l}^{2} + 
R_{ab}l^{a}l^{b}\right) \,.
\eeq  
For situations satisfying the 
null curvature condition, $R_{ab}l^{a}l^{b}\geq 
0$, which can 
be related to the null energy condition, $T_{ab}l^{a}l^{b}\geq 0$, by the Einstein 
equations, $C$ is seen to be negative 
and thus by (\ref{grtraphorvarent}) the area-entropy of the 
future outer trapping horizon is non-decreasing, in which case it is
also spacelike. By  equivalent reasoning an area-entropy law 
can be derived for past inner trapping horizons. In the 
case where a normalisation $l^{a}n_{a}=-1$ is imposed,  the same 
conclusion about area increase can be reached using a  minimum 
principle \cite{Booth:2006bn}.

The existence of the $R_{ab}l^{a}l^{b}$ term in the area 
law gives a direct local relation between the curvature at a 
point and the rate of increase of an area element at that point. 
However, the shear and vorticity terms, although locally defined, 
are related not only to the local properties of the geometry, but 
also to the choice of surface passing through the geometry, {\em 
i.e.}, of the choice of null normals $l^c $ and $n^c$. It is 
perfectly possible, for example, that a portion of the horizon 
can be growing locally in Schwarzschild spacetime, because of the 
non-local influence on the shear and vorticity. This is for 
example what is seen in the merger of vacuum puncture data 
\cite{MOESTA CITE} and is encapsulated in FOTS Property 5 of 
\cite{Booth:2006bn}. In vacuum spacetimes, the shear can only 
increase the area of the trapping horizon and the only way for 
the horizon to shrink is to develop non-zero vorticity.

\subsection{Brans-Dicke theory} 
%**********************************************************************

Brans-Dicke theory is the prototype alternative theory of gravity 
with scalar and tensor modes. The theory was first expressed in a 
frame in which the particle masses remain constant, the 
effective gravitational constant varies from point to point, and 
massive test particles follow timelike geodesics (Jordan or 
string frame). In this frame, the action is given by the 
Lagrangian density \beq \label{bransdicke} {\mathscr{L}} = 
\frac{1}{16\pi}\left( \phi R - 
\frac{\omega}{\phi}\nabla_{a}\phi\nabla^{a}\phi\right) + 
{\mathscr{L}}_{matter} \,. \eeq $\omega$ is the Brans-Dicke 
parameter, not to be confused with the vorticity $\omega_{l}$. 
Variation of this action with respect to the metric gives the 
gravitational field equations \begin{eqnarray} \label{bdeom} && 
G_{ab}\phi = 8\pi T_{ab} + 
\frac{\omega}{\phi}\left(\nabla_{a}\phi\nabla_{b}\phi - 
\frac{1}{2}g_{ab}\nabla_{c}\phi\nabla^{c}\phi \right) + 
\nabla_{a}\nabla_{b}\phi - g_{ab}\nabla_{c}\nabla^{c}\phi 
\,,\nonumber\\ && \end{eqnarray}where 
$T_{ab}$ is the energy-momentum tensor of the matter fields. Contracting the Einstein tensor 
with $l^{a}$ twice for the above yields \beq \label{43} 
R_{ab}l^{a}l^{b} = \frac{8\pi}{\phi} T_{ab}l^{a}l^{b} + 
\frac{\omega}{\phi^{2}}\left(l^{a}\nabla_{a}\phi\right)^{2} + 
\frac{l^{a}l^{b}\nabla_{a}\nabla_{b}\phi}{\phi} \,. \eeq 

In Brans-Dicke theory the horizon-entropy two-form is given by 
$s_{ab}=\phi \varepsilon_{ab}$. The variation of horizon-entropy in 
the outgoing null direction is then \beq \label{entvarBDEH} 
{\cal{L}}_{l}s_{ab} = \left(\theta_{l} + 
\frac{l^{a}\nabla_{a}\phi}{\phi}\right)\phi\varepsilon_{ab} \,. 
\eeq 
Since we require $\phi > 0$ for the horizon-entropy to be 
positive, the term $\left(\theta_{l} + l^{a}\nabla_{a}\phi / 
\phi\right)$ must be positive for the horizon-entropy to be 
increasing for a causal horizon generated by $l^{a}$. The sign of 
the $l^{a}\nabla_{a}\phi$ term though cannot immediately be 
established for a causal horizon. But, by extending a method used 
in \cite{Ford:2000xg}, taking another derivative gives
\beq 
{\cal{L}}_{l}\left(\theta_{l} + 
\frac{l^{a}\nabla_{a}\phi}{\phi}\right) = 
{\cal{L}}_{l}\theta_{l} + \frac{l^{b}\nabla_{b}\left( 
l^{a}\nabla_{a}\phi\right) } {\phi} - 
\frac{1}{\phi^{2}}\left(l^{a}\nabla_{a}\phi\right)^{2} \,.
\eeq  
Using the Raychaudhuri equation (\ref{raychaudhuri}), the 
equations of motion (\ref{bdeom}) and 
\bea 
l^{b}\nabla_{b}\left( l^{a}\nabla_{a}\phi \right) & = & \left( 
l^{b}\nabla_{b}l^{a}\right)\nabla_{a}\phi + l^{a}l^{b}\nabla_{a}\nabla_{b}\phi 
\nonumber \\ & = & \kappa_{l} l^{b}\nabla_{b}\phi + 
l^{a}l^{b}\nabla_{a}\nabla_{b}\phi \,,
\eea  
where $\kappa_{l}$ is 
again a measure of the failure of $l^{a}$ to be affinely 
parameterised, we get 
\bea \label{seconddifflns} 
{\cal{L}}_{l}\left(\theta_{l} + 
\frac{l^{a}\nabla_{a}\phi}{\phi}\right) & = & 
\kappa_{l}\left(\theta_{l} + \frac{l^{a}\nabla_{a}
 \phi}{\phi}\right) - \frac{\theta_{l}^{2}}{2} - \sigma_{l}^{2} + 
\omega_{l}^{2} \nonumber \\ & & - \frac{ \left( \omega + 1 
\right)}{\phi^{2}}\left(l^{a}\nabla_{a}\phi\right)^{2} - 
\frac{8\pi}{\phi}T_{ab}l^{a}l^{b} \,.
\eea 
For a causal horizon 
with $\omega_{l} = 0$, affinely parameterised ($\kappa_{l} = 0$) and 
$\omega + 1 \geq 0$ Brans-Dicke theory, this quantity will be 
negative provided the matter $T_{ab}$ satisfies the null energy 
condition. The condition $\omega_{l} = 0$ is guaranteed because 
the horizon generators are hypersurface orthogonal to the null 
horizon and a  normalisation of the generators can always be 
chosen so that $\kappa_{l} = 0$.  If we then assume that the 
horizon settles down at 
late times to a Killing horizon, such that ${\cal{L}}_{l}s_{ab} = 
0$ at late times, then the term $\left(\theta_{l} + 
\frac{l^{a}\nabla_{a}\phi}{\phi}\right)$ cannot ever be 
negative, because to get from a negative value to zero, its 
derivative must be positive somewhere in between, which is 
excluded by eq.~(\ref{seconddifflns}). Thus the horizon-entropy 
must 
be non-decreasing for a causal horizon, provided it settles down 
at late times to a Killing horizon.
 Note that this requires us to assume that the horizon settles 
down at late times to a Killing horizon, but that this is 
sufficient, we do not need to assume that the horizon forms the 
causal past of future null infinity.  This assumption is not 
needed in the case of Einstein gravity and can be replaced by the 
assumption that the spacetime contains no naked singularities.

In the general case, eq.~(\ref{seconddifflns}) implies that 
if $\theta_{l} +  \frac{l^{a}\nabla_{a}\phi}{\phi}$ were anywhere 
negative on the horizon,  it would reach an infinite value in a 
finite parameter distance.  Thus either $\theta_{l}$ would become 
infinite, implying a focal  point, or 
$ \frac{l^{a}\nabla_{a}\phi}{\phi}$ would become infinite, 
implying a discontinuity in $\phi$. In the former case a focal 
point for the null generators of the horizon is forbidden since 
the generators of the event horizon can have no future 
end-points. We can therefore conclude that if $\phi$ is 
continuous, $\theta_{l} + 
\frac{l^{a}\nabla_{a}\phi}{\phi}$ cannot be negative anywhere on 
the horizon.If the causal horizon is the past causal boundary of some set other than future null infinity then its generators can only have future end-points on the set itself.
 
For a causal horizon, the change of the horizon-entropy cannot be taken 
arbitrarily close to zero in the past if the area remains 
non-zero. If a null surface is initially a Killing horizon with 
zero horizon-entropy change it cannot return to a Killing horizon after a 
perturbation. The equivalent statement in the Einstein case is 
that the expansion of the horizon is always decreasing even 
though it is always positive, so its initial value must be larger 
than any subsequent value. The moment at which the logarithm of 
the horizon-entropy of a causal horizon is changing the most lies 
in the infinite past even though the moment at which the 
horizon-entropy itself  is 
changing the most is not necessarily in the infinite past.

For a trapping horizon we can again use eq.~(\ref{valueC}) but 
now, instead of (\ref{grtraphorvarent}), we have  
\beq 
{\cal{L}}_{r}s_{ab} = \left[ Bl^{c}\nabla_{c}\phi + 
C\left(n^{c}\nabla_{c}\phi + 
\phi\theta_{n}\right)\right]\varepsilon_{ab} \,.
\eeq 
The signs 
of the terms $l^{a}\nabla_{a}\phi$ and $n^{a}\nabla_{a}\phi$ 
cannot be guaranteed from the equations of motion.  Ultimately, 
this is related to the value of $r^{a}\nabla_{a}\phi$ on the 
horizon.  Because of this the horizon-entropy can decrease for a trapping 
horizon \cite{Ford:2000xg}, even in situations where the matter fields obey the null energy condition such as considered in \cite{Scheel:1994yn}.

Because the expansion $\theta_{p}$ of a null congruence with 
tangent $p^{a}$ is related to the variation of the cross-sectional area two-form, 
${\cal{L}}_{p}\varepsilon_{ab} = \theta_{p}\varepsilon_{ab}$, the 
conditions for a future outer trapping horizon (\ref{fothconds}) can be re-written as 
\bea \label{traphorconds} 
 \varepsilon^{ab}{\cal{L}}_{l}\varepsilon_{ab} = 0 \,, \nonumber 
\\
 \varepsilon^{ab}{\cal{L}}_{n}\varepsilon_{ab} < 0 \,,  
\nonumber \\ 
{\cal{L}}_{n}\left(\varepsilon^{ab}{\cal{L}}_{l}\varepsilon_{ab}  
\right) & < 0 \,. 
\eea 
Consider now, instead of future outer trapping horizons, the following conditions: 
\bea \label{modtraphorconds}
 \varepsilon^{ab}{\cal{L}}_{l}s_{ab} = 0 \,, \nonumber \\ 
 \varepsilon^{ab}{\cal{L}}_{n}s_{ab} < 0 \,,  \nonumber \\ 
 {\cal{L}}_{n}\left(\varepsilon^{ab}{\cal{L}}_{l}s_{ab}\right) < 
& 0 \,. 
\eea 
In ordinary Einstein gravity, this would reduce to the 
requirements on the null expansions for a trapping horizon given 
in (\ref{traphorconds}) since, in this case, $s_{ab} = 
\varepsilon_{ab}$ up to a constant factor. But in cases where the 
horizon-entropy is not simply the area, these conditions will in 
general be satisfied at different locations of the spacetime. In 
Brans-Dicke theory the horizon-entropy two-form is just $s_{ab} = \phi 
\varepsilon_{ab}$ in which case the first condition is satisfied 
where $\phi\theta_{l} + l^{a}\nabla_{a}\phi = 0$ and the second 
condition when $\phi\theta_{n} + n^{a}\nabla_{a}\phi < 0$.

The variation of the horizon-entropy two-form is now \bea 
\label{genareavar} {\cal{L}}_{r}s_{ab} & = & B{\cal{L}}_{l}s_{ab} 
+ C{\cal{L}}_{n}s_{ab} \nonumber \\ & = & C{\cal{L}}_{n}s_{ab} 
\nonumber \\ & = & C\left(\phi\theta_{n} + 
n^{c}\nabla_{c}\phi\right)\varepsilon_{ab}. \eea The first term 
on the right hand side of the first line is now zero by 
assumption. The term $\left(\phi\theta_{n} + 
n^{c}\nabla_{c}\phi\right)$ is negative by assumption and so the 
sign of the change in horizon-entropy along the horizon is given by the sign 
of $C$ again. If $C$ is negative, the horizon is spacelike and 
the horizon-entropy increases.

It is possible to determine the sign of $C$ by a similar argument 
used for trapping horizons. Since we require the tangent $r^{a}$ 
to generate evolution along a horizon on 
which ${\cal{L}}_{l}s_{ab}=0$, we have \beq C \label{valuec} = 
-\frac{B{\cal{L}}_{l}\left(\varepsilon^{ab}{\cal{L}}_{l}s_{ab}\right) 
}{{\cal{L}}_{n}\left(\varepsilon^{cd}{\cal{L}}_{l}s_{cd}\right)} 
\,. \eeq With the sign of $B$ assumed positive, setting the 
orientation of $r^{a}$, and 
${\cal{L}}_{n}\left(\varepsilon^{cd}{\cal{L}}_{l}s_{cd}\right)$ 
negative by assumption on the horizon, whether the horizon-entropy is increasing or not is just determined by the sign of the term 
${\cal{L}}_{l}\left(\varepsilon^{ab}{\cal{L}}_{l}s_{ab}\right)$: 
\begin{equation} \mbox{sign}\left( 
\varepsilon^{ab}{\cal{L}}_{r}s_{ab} \right)=- \mbox{sign}\left( 
{\cal{L}}_{l}\left(\varepsilon^{ab}{\cal{L}}_{l}s_{ab}\right) 
\right) \,. \end{equation} Using eqs.~(\ref{raychaudhuri}) and 
(\ref{43}), we obtain \beq \label{entvaryconf} 
{\cal{L}}_{l}\left(\varepsilon^{ab}{\cal{L}}_{l}s_{ab}\right) = 
2\phi\left( - \frac{1}{2} \, \theta_{l}^{2} - \sigma_l^{2} - 
\frac{\omega + 1}{\phi^{2}}\left( l^{a}\nabla_{a}\phi \right) 
^{2} - \frac{8\pi}{\phi} T_{ab}l^{a}l^{b} \right) \,. \eeq For 
$\omega + 1 \geq 0$ and matter obeying the null energy condition 
$ T_{ab}l^{a}l^{b} \geq 0$, the horizon-entropy is guaranteed 
to be non-decreasing along surfaces satisfying 
(\ref{modtraphorconds}).\footnote{In fact, because the 
conditions~(\ref{modtraphorconds}) are satisfied in Brans-Dicke 
theory by surfaces satisfying $\theta_{l} = - l^{a}\nabla_{a}\phi 
/\phi$, the $\theta_{l}$ term can be eliminated 
in eq.~(\ref{entvaryconf}) and the condition on the Brans-Dicke 
parameter $\omega$ becomes $\omega > -3/2$.} The condition 
in (\ref{entvaryconf}) is very similar to (\ref{seconddifflns}) 
for a causal horizon, except now the term involving $\kappa_{l}$ in (\ref{seconddifflns}) 
vanishes on a horizon satisfying (\ref{modtraphorconds}) anyway, 
and the horizon-entropy is guaranteed to increase without an 
assumption that it settles down to a future Killing horizon. We 
remind the reader that the location of surfaces for which these conditions 
hold will in general be different from causal horizons. Surfaces satisfying (\ref{modtraphorconds}) will be spacelike for 
positive energy.

The similarity of (\ref{entvaryconf}) to (\ref{seconddifflns}) is 
not surprising since we have \beq 
{\cal{L}}_{l}\left(\varepsilon^{ab}{\cal{L}}_{l}s_{ab}\right) = 
2{\cal{L}}_{l}\phi\left(\frac{{\cal{L}}_{l}\phi}{\phi} + 
\theta_{l}\right) + 
2\phi{\cal{L}}_{l}\left(\frac{{\cal{L}}_{l}\phi}{\phi} + 
\theta_{l}\right) \,. \eeq The first term is zero by the 
assumption $\varepsilon^{ab}{\cal{L}}_{l}s_{ab} = 0$ and the 
second term is just eq.~(\ref{seconddifflns}).  The right hand 
side of (\ref{entvaryconf}) is used in the first variation of 
the horizon-entropy for quasi-local  horizons, through eqs.~(\ref{genareavar}) and~(\ref{valuec}), but in the 
second variation of the horizon-entropy for causal 
horizons,  through eqs.~(\ref{entvarBDEH}) 
and~(\ref{seconddifflns}).  If the right hand side of 
(\ref{entvaryconf})  ever becomes positive then the 
horizon-entropy of the  quasi-local horizons will immediately 
start to decrease, but  the change of horizon-entropy of a causal horizon 
may still increase because  in this case it only influences the 
second variation of the horizon-entropy.  

In the case where we impose a cross-normalisation  
$l^{a}n_{a}=-1$ as is done in~\cite{Booth:2006bn}, we do not 
have complete freedom to rescale  $l^{a}$ and $n^{a}$ so that $B$ 
and $C$ in (\ref{rwithBandC}) are  constant. In this case, the 
$r^{a}$ variation, $\delta_{r}$, as  defined in 
\cite{Booth:2006bn}, is not equivalent  to the Lie derivative with 
respect to $r^{a}$ for terms such as $ \theta_{l}$ that depend 
not just on the spacetime point but also  on the choice of 
two-surface for which they are defined. The  variation of 
$\theta_{l} + {\cal{L}}_{l}\phi/\phi$ however  splits into a 
variation of $\theta_{l}$ and a part that is  equivalent to the 
Lie derivative because $\phi$ is a globally  defined scalar 
field. In this case a maximum principle can  still be invoked as 
in \cite{Booth:2006bn} since the variation becomes
\bea \noindent \delta_{r}\left(\theta_{l} +  
\frac{{\cal{L}}_{l}\phi}{\phi}\right) = & & \kappa_{r}\theta_{l} + 
\d^{2}C - 2\tilde{\omega}^{a}\d_{a}C \nonumber \\ & & + B{\cal{L}}_{l}\left(\theta_{l} + \frac{{\cal{L}}_{l}\phi}{\phi}\right) + C{\cal{L}}_{n}\left(\theta_{l} + \frac{{\cal{L}}_{l}\phi}{\phi}\right) \,, 
\eea
with notation adapted from \cite{Booth:2006bn}. The term 
involving $B$ is once again (\ref{entvaryconf}). In the case 
where this term is negative and $l^{a}$ and $n^{a}$ are both 
derived from a double null foliation so that $\kappa_{r}$ 
vanishes, a maximum principle can be applied (see 
\cite{Booth:2006bn} for further details) to conclude that $C$ is 
either constant or everywhere negative and the horizon-entropy 
is non-decreasing.

\subsection{Scalar-tensor and $f(R)$ gravity} 
%****************************************************************

The scalar-tensor generalizations of Brans-Dicke theory, 
described by the action 
\begin{equation}\label{STaction} 
S_{ST}=  \int \d^4x \sqrt{-g} \, \left[ \frac{ F(\phi)R}{16\pi} 
-\frac{\omega(\phi)}{\phi} \, \nabla^a\phi \nabla_a\phi -V(\phi) 
\right] +S_{matter} \,, 
\end{equation} 
where the Brans-Dicke  coupling $\omega$ becomes a function of 
$\phi$ and a scalar field  potential $V(\phi)$ is introduced, can 
be discussed in the same  way as Brans-Dicke theory. One can 
consider a new Brans-Dicke  field $\psi \equiv F(\phi)$ provided 
that the function $ F(\phi)$ 
admits a regular inverse $F^{-1}$ (this is not always the case in 
the literature, in which $ F(\phi)$ is sometimes found in the 
form  of a series of even powers of $\phi$ \cite{LiddleWands92, 
TorresVucetich96}, but specific choices in the literature are 
motivated by mathematical, not physical considerations, {\em 
i.e.}, by the fact that they allow certain calculations to be 
performed). Then the action~(\ref{STaction}) can be recast in the 
form 
\begin{equation} 
S_{ST}= \int \d^4x \sqrt{-g} \, \left[  \frac{\psi R}{16\pi} 
-\frac{\bar{\omega}(\psi)}{\psi} \,  \nabla^a\psi\nabla_a\psi 
-U(\psi) \right] +S_{matter} \,; 
\end{equation} 
therefore, we limit ourselves to consider the 
action~(\ref{STaction}) with $ F(\phi)=\phi$, which yields the 
field equations 
\begin{eqnarray} 
G_{ab} = & & \frac{8\pi}{\phi}\, 
T_{ab} +\frac{ \omega(\phi)}{\phi^2} \left( \nabla_a\phi 
\nabla_b\phi -\frac{1}{2} \, g_{ab} \, \nabla^c\phi \nabla_c\phi 
\right) \label{37} \nonumber \\ & & + \frac{1}{\phi} \left( 
\nabla_a \nabla_b \phi -g_{ab} \Box \phi\right)
- \frac{V(\phi)}{2\phi}\, g_{ab} \,, \label{ST1}\\ &&\nonumber\\ 
  \nabla^{a}\nabla_{a}\phi & = & \frac{1}{2\omega+3} \left( 8\pi 
  T -\frac{\d\omega}{\d\phi} \, \nabla^c\phi \nabla_c\phi +\phi \, 
  \frac{\d V}{\d\phi} -2V \right) \,. \label{ST2} 
\end{eqnarray} 
The  discussion of horizons in scalar-tensor gravity remains the 
  same as in Brans-Dicke theory because, by contracting 
  eq.~(\ref{ST1}) twice with the null vector $l^a$, one obtains 
  again eq.~(\ref{43}) (now with $\omega$ dependent on $\phi$). 
  Since the horizon-entropy is again $S=\phi A$, one finds 
  again eqs.~(\ref{seconddifflns}) and (\ref{entvaryconf}).

Metric modified (or $f(R)$) gravity, described by the action 
\begin{equation} S_{MG}=\frac{1}{16\pi}\int \d^4x \sqrt{-g} \, 
f(R)+S_{matter} \end{equation} is equivalent to a Brans-Dicke 
theory with $\omega=0$ and a potential \cite{f(R)reviews}. In 
fact, setting 
\begin{equation} 
\phi =f'(R) \,, \;\;\;\;\;\; 
V(\phi)=\phi R(\phi)-f(R(\phi)) 
\end{equation} 
leads to the 
equivalent action \cite{f(R)reviews} 
\begin{equation} 
S'_{MG} 
=\frac{1}{16\pi}\int \d^4x \sqrt{-g} \left[ \phi R -V(\phi) 
\right]+S_{matter} 
\end{equation} 
(similarly, Palatini $f(R)$ 
gravity can be recast as an $\omega=-3/2$ Brans-Dicke theory with 
a potential, but we will not consider it here because of its 
well-known problems \cite{f(R)reviews}).  Since the potential 
$V(\phi)$ does not give contributions upon double
contraction of eq.~(\ref{37}) with the null vector $l^a$, the 
considerations on horizons made for Brans-Dicke theory can be 
immediately extended to $f(R)$ gravity.

%***************************************************************** 
\section{Horizons under conformal transformations} 
%***************************************************************** 
\label{sec:conformaltransformations}

A conformal transformation of the metric will, in general, change 
the areas of spacelike two-surfaces. This in turn will change the 
location of the trapping horizons given by the above conditions 
(\ref{traphorconds}). The conformal transformation relates two 
different conformal frames if the metric is scaled by a conformal 
factor that can vary with spacetime point \beq 
\label{conformaltrans} g_{ab} \rightarrow \tilde{g}_{ab} = 
W(x)g_{ab} \,. \eeq The geometric expansion of a null vector 
$l^{a}$ in any frame is given by \beq \label{expansion} 
\theta_{l} = q^{ab}\nabla_{a}l_{b} = \left( g^{ab} + 
\frac{l^{a}n^{b}}{\left( -n^{c}l^{d}g_{cd}\right)} + 
\frac{n^{a}l^{b}}{\left( 
-n^{c}l^{d}g_{cd}\right)}\right)\nabla_{a}l_{b} \,, \eeq where 
${q_a}^b $ is a projection tensor onto the two-dimensional 
spacelike surface to which $l^{a}$ and $n^{a}$ are normal. (If 
$l^{a}$ is defined as globally null, then the third term on the 
right hand side vanishes identically.) This result holds for a 
Lorentzian signature manifold independently of whether the 
Einstein equations hold. In general, there is freedom to rescale 
null vectors even without rescaling the metric. The vanishing of 
the expansion does not depend on a pure rescaling of the null 
vector $l^{a} \rightarrow Wl^{a}$, although its value does since 
under this rescaling we have \beq \theta_{l} \rightarrow 
W\theta_{l} \,. \eeq Under a conformal transformation of the 
form~(\ref{conformaltrans}) we have 
${\tilde{g}}^{ab}=W^{-1}g^{ab}$ and $ q^{ab} \rightarrow 
\tilde{q}^{ab}= W^{-1}q^{ab}$. We can fix the normalization of 
$l^{a}$ by requiring ${\tilde{l}}^{a} = l^{a}$ with 
${\tilde{l}}_{a} = Wl_{a}$ and thus \beq 
{\tilde{\nabla}}_{a}{\tilde{l}}_{b} = W\nabla_{a}l_{b} + 
l_{b}\nabla_{a}W - \frac{1}{2} \left(l_{a}\nabla_{b}W + 
l_{b}\nabla_{a}W - g_{ab}l^{c}\nabla_{c}W\right) \,, \eeq 
therefore \beq \label{thetaltrans} {\tilde{\theta}}_{l} = 
\theta_{l} + \frac{l^{a}\nabla_{a}W}{W} \,. \eeq The vanishing of 
$\theta_{l}$ for a given surface is therefore not necessarily 
invariant under a conformal transformation. And thus the location 
of a marginally outer trapped surface satisfying $\theta_{l}=0$ 
is not necessarily invariant. This is despite the fact that the 
conformal transformation does not change the coordinates of a 
given spacetime event nor the path of null rays. The location of 
the event horizon, for example, is unchanged. In one frame the 
solution of $\theta_{l}=0$ may lie inside the event horizon and 
in another frame outside, as discussed in \cite{Scheel:1994yn}.

The vanishing of the expansion is equivalent to the statement 
that the area is unchanged under infinitesimal translations along 
$l^{a}$ via the relation ${\cal{L}}_{l} \, \varepsilon_{ab} = 
\theta_{l} \, \varepsilon_{ab}$. Since the conformal factor 
changes how areas are measured, this criterion no longer selects 
the same horizon in the two frames. The area two-form changes as 
$\varepsilon_{ab} \rightarrow \tilde{\varepsilon}_{ab}= W 
\varepsilon_{ab}$. The condition that the Lie derivative of this 
``conformally transformed area'' be zero is \beq 
{\cal{L}}_{l}\left( W \varepsilon_{ab}\right ) = \left( 
\theta_{l} + \frac{{\cal{L}}_{l}W}{W}\right)W\varepsilon_{ab} = 0 
\,. \eeq This rule is the same as the transformation in 
(\ref{thetaltrans}). This also makes clear what lies behind the relations 
(\ref{modtraphorconds}). In the Einstein frame the horizon-entropy is 
just one quarter of the area but the area is modified in other 
conformal frames, leading to a modification of the relationship 
between horizon-entropy and area. The formula~(\ref{waldentropy}) 
gives an explicit way of calculating this new horizon-entropy in the new 
frame and for the class of theories we have investigated the 
entropy is invariant.  For example, the gravitational 
sector in the Einstein frame has the familiar Einstein-Hilbert 
form of the action \beq S_{action} = \int\d^{4}x\sqrt{-g}R \,. 
\eeq By~(\ref{waldentropy}), the horizon-entropy is $A/4$ in the 
Einstein frame. In another frame obtained by $g_{ab}\rightarrow 
\tilde{g}_{ab} = Wg_{ab}$, the same action will take the form \beq 
S_{action} = 
\int\d^{4}x\sqrt{-\tilde{g}}\left(\frac{\tilde{R}}{W}
- \frac{9}{2}\frac{\tilde{g}^{ab}}{W^{3}} \, 
  \tilde{\nabla}_{a}W\tilde{\nabla}_{b}W + 3 \, 
  \frac{\tilde{g}^{ab}}{W^{2}}\tilde{\nabla}_{a} 
  \tilde{\nabla}_{b}W\right) \,, \eeq and the horizon-entropy, 
  with $W$ constant on the horizon, will be $\tilde{A}/{4W}$. 
  But, since the areas are related by $\tilde{A} = AW$, the 
  horizon-entropy will still take the same numerical value in the new 
  frame (the equality between entropies in the Einstein and 
  the Jordan frames extends to all theories with action $ \int 
  d^4x \, \sqrt{-g} \, f\left( g_{ab}, R_{ab}, \phi, \nabla_c 
  \phi \right)$ \cite{KogaMaeda}). As long as the horizon-entropy 
  transforms in the same way as the metric under a conformal 
  transformation, the conditions (\ref{modtraphorconds}) will 
  give rise to a surface whose location is invariant and for 
  which one can derive a horizon-entropy increase law, exactly as 
  one can derive an area increase law for trapping horizons in 
  the Einstein frame.

The issues that occur can be illustrated with a few examples. One 
of the cases considered in \cite{Deruelle:2010ht} is the 
Schwarzschild spacetime under a conformal transformation with 
conformal factor $W=\Delta^{-1}$. Then the ``veiled general 
relativity'' spacetime becomes \beq
 \d \tilde{s}^{2} = -\d t^{2} + \frac{\d r^{2}}{\Delta^{2}} + 
\frac{r^{2}}{\Delta} \, \d\Omega_2^{2} \,, \eeq where $\Delta= 
1-2M/r$ and $d\Omega_2^2=d\theta^2+\sin^2 \theta \, d\varphi^2$ 
is the line element on the unit two-sphere. Like the usual form 
of the Schwarzschild metric in Schwarzschild coordinates, this 
metric is valid everywhere in the region $r>2M$. The radial null 
vectors in this transformed metric have components 
\begin{equation} \label{ddd} l^{\mu}=\left( 1, \Delta, 0,0 
\right) \,, \;\;\;\;\;\;\;\; n^{\mu}=\left( 1, - \Delta, 0,0 
\right) \,, \end{equation} and their expansions are, using 
eq.~(\ref{expansion}), \beq \tilde{\theta}_{l} = 
\frac{2}{r^{2}}\left(r-3M\right) \,, \;\;\;\;\;\;\;\; 
\tilde{\theta}_{n} = -\frac{2}{r^{2}}\left(r-3M\right) \,. \eeq 
The marginally outer trapped surfaces, where $\theta_{l}$ 
vanishes, are now found at $r=3M$ instead of $r=2M$. But these 
surfaces do not form a trapping horizon because $\theta_{n}=0$ 
here too. The constant $r$ tube at $r=3M$ is timelike, but its 
area is not decreasing because $\theta_{n} = \theta_{l} = 0$.  
There are no true spherically symmetric trapping horizons in this 
metric. In fact, in this metric there are not even any 
spherically symmetric trapped surfaces, because nowhere do we 
have $ \tilde{\theta}_{l}\tilde{\theta}_{n}>0 $.

This result can be generalised by considering conformal factors 
of the form $W(x) = \Delta^{n}$, in which case the null vectors 
are given again by eq.~(\ref{ddd}) and their expansions become 
\beq \tilde{\theta}_{l} = \frac{2}{r^{2}}\left[ r-M(2-n)\right] 
\,, \;\;\;\;\;\;\;\;
 \tilde{\theta}_{n} = -\frac{2}{r^{2}}\left[ r-M(2-n)\right] \,. 
\eeq The marginally outer trapped surfaces can be conformally 
transformed to any $r$ by a suitable choice of $n$. But since the 
area of the spherically symmetric two-spheres is now $A = 
4\pi\Delta^{n}r^{2}$, the horizon-entropy is $S = A/W = 
\Delta^{-n} A = 4\pi r^{2}$, and the conditions 
(\ref{modtraphorconds}) just give back trivially the location of 
the usual horizon, $r=2M$.

Similar things can occur with coordinates and conformal factors 
that are perfectly regular on the horizon. For example, in 
Kerr-Schild coordinates the Schwarzschild metric takes the form 
\beq \d s^{2} = -\Delta\d t^{2} + 2(1-\Delta)\d t\d r + 
(2-\Delta)\d r^{2} + r^{2}\d\Omega_2^{2} \,. \eeq In these 
coordinates the radial null vectors have components \beq l^{\mu} 
= \left( 1-\frac{\Delta}{2}, \frac{\Delta}{2}, 0 ,0 \right) \,, 
\;\;\;\;\;\;\;\; n^{\mu} = \left( 1, -1, 0 ,0 \right) \,, \eeq 
where $n^{\mu}$ is affinely parameterised, {\em i.e.}, 
$n^a\nabla_a n^b=0$. In this frame the expansions are, as 
expected, \beq \theta_{l} = \frac{\Delta}{r}\,, \hspace{3cm} 
\theta_{n} = -\frac{2}{r} \,. \eeq If we choose a conformal 
factor of the form $W(x) = e^{-\lambda t^2}$ we find \bea 
\tilde{\theta}_{l} & = & \frac{\lambda rt\Delta + \Delta - 
2\lambda rt}{r} \,, \nonumber \\ \tilde{\theta}_{n} & = & 
-\frac{2\left( \lambda rt + 1\right)e^{\lambda t^{2}}}{r} \,. 
\eea Setting $\tilde{\theta}_{l}$ equal to zero and expanding for 
$\lambda t \ll 1/M$ gives \beq r = 2M + 8\lambda t M^{2} + 
{\cal{O}}(\lambda^{2}) \,. 
\eeq 
 In this limit the trapping 
horizon is close, but not equal, to the $r=2M$ surface, but it is 
now also spacelike. The surface $r=2M$ is still null and is still 
the location of the event horizon and again is given simply by 
the conditions (\ref{modtraphorconds}). The physical horizon 
is located by~(\ref{modtraphorconds}), not 
by~(\ref{traphorconds}).

It has been observed in numerical simulations of black hole 
collapse in Brans-Dicke theory that the trapping horizon can 
appear outside the event horizon \cite{Scheel:1994yn, 
BDcollapse}.  This possibility occurs despite the fact that the 
Jordan frame of Brans-Dicke theory (\ref{bransdicke}) can be 
related via a conformal transformation to Einstein theory with a 
scalar field that obeys the null energy condition. In the 
Einstein frame the trapping horizon appears exclusively inside 
the event horizon, in accordance with a theorem of Hawking and 
Ellis \cite{Hawking:1973uf}.

Two related issues are involved here. First, unlike the event 
horizon, the location of the trapping horizon changes under a 
conformal transformation. Second, the null energy condition is 
not necessarily equivalent to the null curvature condition.  
(This condition is called the null convergence condition 
in~\cite{Hawking:1973uf}.) The 
trapping horizon can appear outside the event horizon in the 
string frame, even if the null energy condition is satisfied, 
because the Einstein equations do not hold in this frame 
\cite{Scheel:1994yn, BDcollapse}.

The proof that the apparent horizon cannot lie outside the event 
horizon (the apparent horizon theorem \cite{Hawking:1973uf}) is 
purely geometric and relies only on the validity of the null 
Raychaudhuri equation (\ref{raychaudhuri}) and the geometrical 
condition $R_{ab}l^{a}l^{b} \geq 0$, the null curvature 
condition.  Even if the matter obeys the null energy condition, the sign of the last term in (\ref{43}) 
can be negative and, therefore, we may have a violation of the 
null curvature condition. This is in fact what happens for the 
surfaces found in \cite{Scheel:1994yn, BDcollapse}. Brans-Dicke 
theory can be recast in the Einstein frame via the conformal 
transformation \beq \label{bdconftrans} g_{ab} \rightarrow 
{\tilde{g}}_{ab} = \phi \, g_{ab} \,, \;\;\;\;\;\; \phi 
\rightarrow \tilde{\phi} \;\; \mbox{with} \;\; 
\d\tilde{\phi}=\sqrt{ \frac{2\omega+3}{16\pi} }\, 
\frac{\d\phi}{\phi} \,. \eeq Here $\phi >0 $ in order to 
guarantee that the effective gravitational coupling $G_{eff}\sim 
\phi^{-1}$ remains positive.  In the Einstein frame the null 
tangent vectors are unchanged, ${\tilde{l}}^{a} = l^{a}$, and 
they are null with respect to the ``new'' metric $\tilde{g}_{ab}$ 
as well as the old one $g_{ab}$. In the Einstein frame the 
gravitational field equations are 
\beq 
{\tilde{G}}_{ab} = 8\pi 
\left( {\tilde{T}}_{ab} + {\tilde{\nabla}}_{a} \tilde{\phi} 
{\tilde{\nabla}}_{b} \tilde{\phi}
-  \frac{1}{2}{\tilde{g}}_{ab}  
\, \tilde{g}^{cd}{\tilde{\nabla}}_{c} 
  \tilde{\phi}{\tilde{ \nabla}}_{d} \tilde{\phi} \right) \,, 
\eeq 
where $\tilde{T}_{ab}\equiv T_{ab}/\phi^2$ and thus 
\beq 
  \tilde{g}^{ac}\tilde{g}^{bd} {\tilde{R}}_{ab} l_{c}l_{b} = 8\pi 
  \tilde{g}^{ac}\tilde{g}^{bd} {\tilde{T}}_{ab}l_{c}l_{d} + 
  \left( \tilde{g}^{ab} l_{a}{\tilde{\nabla}}_{b} \tilde{\phi} 
  \right)^{2} \,. 
\eeq 
Provided the matter obeys the null energy condition and $\omega
> 0$, the geometry will also obey the null curvature condition in
the Einstein frame. As we have seen, quasi-local horizons satisfying 
(\ref{modtraphorconds}) cannot appear outside the event horizon 
in any conformal frame if they are located entirely inside the 
event horizon in the Einstein frame.

%***************************************************************** 
\section{Cosmological horizons} 
%***************************************************************** 
\label{sec:cosmologicalhorizons}

The problem of locating the trapping horizons following a 
conformal transformation appears not only in black hole 
spacetimes but also in cosmology in alternative gravity 
(especially scalar-tensor and $f(R)$ theories, which can be 
formally reduced to Einstein gravity plus non-minimally coupled 
scalars by a conformal transformation). The line element of a 
spatially homogeneous and isotropic 
Friedmann-Lemaitre-Robertson-Walker (FLRW) metric is commonly 
written in comoving coordinates as 
\beq \label{FRWmetric} 
\d s^{2} = -\d t^{2} + a(t)^{2}\left(\frac{\d r^{2}}{1-kr^{2}} + 
r^{2}\d\theta^{2} + r^{2}\sin^{2}\theta\d\phi^2\right) \,. 
\eeq 
In the spatially flat case $k=0$, we have radial null vectors 
with components \beq l^{\mu} = \left( 1, 1/a(t), 0 ,0 \right) \,, 
\;\;\;\;\;\;\;\; n^{\mu} = \left( 1, -1/a(t), 0 ,0 \right) \,, 
\eeq which have expansions \beq \theta_{l} = 
\frac{2\left(\dot{a}r+1\right)}{ar} \,, \;\;\;\;\;\;\;\; 
\theta_{n} = \frac{2\left(\dot{a}r-1\right)}{ar} \,. \eeq We see 
that $\theta_{n} = 0$ when the comoving radius is $r=1/\dot{a}$ 
and the physical radius is $r_{physical}= ar= H^{-1}$, the usual 
Hubble radius.\footnote{The physical horizon radius can also be 
obtained by rewriting the line element in the form 
$ds^2=-\left(1-H^2R^2 \right)dt^2+ \left(1-H^2R^2 \right)^{-1} 
dR^2 +R^2 \left( d\theta^2 +\sin^2 \theta \, d\varphi^2 \right)$, 
where $R=ar$, and finding the root of 
$g^{11}=0$.} The expansion $\theta_{l}$ is everywhere positive 
for $r>0$ and $\dot{a}>0$. We can also calculate the variation of 
the expansions with respect to the null directions \beq 
{\cal{L}}_{l}\theta_{n} = \frac{2\left(-\dot{a}^{2}r^{2} + 
r\dot{a} + ar^{2}\ddot{a}+1\right)}{a^{2}r^{2}} \,, \eeq which 
equals \beq {\cal{L}}_{l}\theta_{n} = 2\left( 
\frac{\dot{a}}{a}\right)^{2}\left(1 + 
\frac{\ddot{a}a}{\dot{a}^{2}}\right) =2H^{2}\left(1-q\right) \eeq 
on the horizon $r=1/\dot{a}$. The expansion is accelerating if 
$q<0$. For de Sitter space, $q=-1$. Thus, for de Sitter space we 
have a Past Inner Trapping Horizon (PITH) by the classification 
of \cite{Hayward:1993wb}. This is reasonable because the region 
around $r=0$ is not trapped ($\theta_{l}\theta_{n} < 0$), so it 
should be an inner horizon, and a past horizon because beyond the 
horizon everything must move outwards, nothing can fall back. The 
components of the normal $N_{a}$ to the surface $r=1/\dot{a}$ are 
\beq N_{\mu} = \left( \frac{\ddot{a}}{\dot{a}^{2}}, 1, 0, 0 
\right) \,. 
\eeq 
The  norm squared of this normal is 
\beq 
N^{a}N_{a} = -\left(\frac{\ddot{a}}{\dot{a}^{2}}\right)^{2}
+ \frac{1}{a^{2}} \,\, = \,\, \frac{1}{a^{2}}\left(1-q^{2}\right) \,. 
\eeq 
For  de Sitter spacetime the horizons are null. In the more 
general $k \neq 0 $ case, we have 
\beq 
l^{\mu} = \left( 1,  \frac{\sqrt{1-kr^{2}}}{a(t)}, 0 ,0 \right) 
\,, \;\;\;\;\;\;\;\; 
n^{\mu} = \left( 1, -\frac{\sqrt{1-kr^{2}}}{a(t)}, 0 ,0 \right)  
\,, 
\eeq 
which have expansions 
\beq 
\theta_{l} = \frac{2\left(\dot{a}r+\sqrt{1-kr^{2}}\right)}{ar} \,, 
  \;\;\;\;\;\;\;\; \theta_{n} = 
  \frac{2\left(\dot{a}r-\sqrt{1-kr^{2}}\right)}{ar} \,. 
\eeq 
Here  we have horizons at 
\beq 
r = \frac{1}{\sqrt{\dot{a}^{2}+ k}} \,. 
\eeq 
(Note that in general relativity, due to the  Hamiltonian 
constraint $H^2=8\pi G \rho/3 -k/a^2$, this radius  is always 
real if the energy density $\rho$ is  positive-definite.)

By transforming to the conformal time $\eta$ defined by 
$a(\eta)\d\eta = \d t$, the metric (\ref{FRWmetric}) can be cast 
in the form 
\beq \label{manifestflrw} 
\d s^{2} = a^2(\eta) 
\left[ -\d\eta^{2} + \frac{\d r^{2}}{1-kr^{2}} + 
r^{2} \left( \d\theta^{2}  + \sin^{2}\theta\d\phi^2\right) 
\right] \,. 
\eeq 
This line element is  manifestly 
conformally flat for the spatially flat case $k=0$ but, because 
the Weyl tensor vanishes in all cases, all FLRW metrics are 
conformally flat, even for $k\neq 0$. So we relate the metrics by 
a conformal transformation \beq g_{\mathrm{FLRW}} = 
a^{2}g_{\mathrm{flat}} \,. \eeq so that $ g^{flat}_{ab}= 
\tilde{g}_{ab}=W \, g_{ab}$ with $W=1/a^2(\eta)$.

From the point of view of ``veiled'' general relativity 
\cite{Deruelle:2010ht}, an expanding FLRW universe is physically 
equivalent to its flat conformal cousin. This equivalence is 
apparently surprising and it helps to consider the variation of 
units of length $l_u$, time $t_u$, mass $m_u$, and of the derived 
units here \cite{Dicke, Faraoni:2006fx}. In the flat 
``veiled frame'' spacetime these units are not fixed but scale 
as $\tilde{l}_u \sim W^{1/2} l_u = a^{-1} l_u $, $ \tilde{t}_u 
\sim W^{1/2}  t_u = a^{-1} t_u$, $ \tilde{m}_u \sim W^{-1/2} m_u 
= a m_u $ (the scaling of derived units is argued by straightforward dimensional 
considerations). Despite appearances, gravity is still present in 
this space and it acts by shrinking the units $\tilde{l}_u$ and $ 
\tilde{t}_u$ instead of making the universe expand as in the 
original FLRW space. Thus, we do not have a genuine Minkowski 
space, but one with time-dependent units, a fact that must be 
kept in mind at all times.  Actual measurements are always made 
with respect to a unit scale.  A given time interval, for 
example, is recorded by dividing  it up into blocks of the time 
unit $t_{u}$.\footnote{In FLRW space the spatial homogeneity 
and isotropy select a preferred family of observers, the comoving 
observers  who see the cosmic microwave background homogeneous 
and isotropic around them (apart from small temperature 
fluctuations $\delta T/T \sim 5 \cdot 10^{-5}$).  The comoving 
time $t$ is the proper time of these observers, hence it is  a 
geometrically and physically preferred notion of time.} There 
is no physical difference 
between a static space with all lengths and times shrinking, or 
an expanding FLRW space with fixed units. For example, in this 
static space in the frame with varying units, there is 
cosmological redshift (which is obviously absent in a genuine 
Minkowski space with fixed units), caused by the fact that the 
unit of length $\tilde{l}_u$ assumes different values at the 
different instants of emission and observation of a light signal. It is instructive to derive this redshift in the flat space 
with line element  
\begin{equation}\label{star} 
\d\tilde{s}^2=-\d\eta^2 +\d r^2 
+r^2 \left( \d\theta^2 +\sin^2 \theta \, \d\varphi^2 \right) \,.
\end{equation}
To keep track of the variation of units in the ``veiled frame'', 
divide by the unit of length squared in this frame and use the 
fact that $ \tilde{l}_u=\tilde{t}_u$  (this is merely a choice of 
units so that everywhere the measured speed of light is $1$),
\begin{equation} 
\frac{\d\tilde{s}^2}{ \tilde{l}_u^2} = - \, \frac{\d\eta^2}{ 
\tilde{t}_u^2 }  + \frac{\d r^2}{ \tilde{l}_u^2}  
+ \frac{r^2}{\tilde{l}_u^2}  \left(\d\theta^2 +\sin^2 \theta \, 
\d\varphi^2 \right) \,.
\end{equation}
The time interval between two equal intervals of $\eta$, say 
$\int^{\eta_{2}}_{\eta_{1}}\d\eta$ and 
$\int^{\eta_{4}}_{\eta_{3}}\d\eta$ such that $\eta_{2}-\eta_{1} 
= \eta_{4}-\eta_{3}$, will not be measured by a comoving 
observer to be equal amounts of time, not because the $\eta$ 
intervals are different, but because the units with which they 
are compared are changing with time. In the first term on the 
right hand side the square of the ratio 
$\d\eta/\tilde{t}_u$ appears, but one must compare $t$-time 
intervals with  the unit $\tilde{t}_u$ and $\eta$-time intervals 
with the unit $ \tilde{\eta}_u$, hence we convert $\eta$ to $t$ 
using  an ordinary coordinate transformation (not a 
transformation of units) $\d t=a\d\eta$, obtaining
\begin{equation} 
\frac{\d\tilde{s}^2}{ \tilde{l}_u^2} = - \, \frac{\d t^2}{ 
a^2 \tilde{t}_u^2 }  + \frac{\d r^2}{ \tilde{l}_u^2}  
+ \frac{r^2}{\tilde{l}_u^2}  \left(\d\theta^2 +\sin^2 \theta \, 
\d\varphi^2 \right) \,.
\end{equation}
Equal intervals of the $t$ coordinate will be measured as equal 
time intervals with respect to the fundamental unit scale. Even 
though $a$ is a function of $t$ this line element is still 
manifestly flat. Consider now a light ray emitted at radius 
$r_e$ at time $t_e$, 
which propagates radially and is received by an observer at 
$r=0$ at time $t_o$. Setting $\d\tilde{s}^2=0$ and 
$\d\theta=\d\varphi=0$ for radial null geodesics\footnote{A null 
geodesic ($\d s^2=0$) in the original frame corresponds to a null 
geodesic ($\d\tilde{s}^2=W\d s^2=0$) in the ``veiled frame'' (it is 
not 
so for {\em timelike} geodesics).} one obtains  
$ \frac{\d t}{a(t)}= \pm \d r $, where the negative sign must 
be chosen for rays propagating from $r_e$ to $r=0$ along the 
direction of decreasing $r$.  Integrating between 
emission and observation yields  
\begin{equation} 
\int_{t_e}^{t_o} 
\frac{\d t}{a(t)} =- \int_{r_e}^{0}\d r \,. 
\label{radialintegral1} 
\end{equation} 
Consider now a second  pulse emitted at $r_e$ at time 
$t_e+\delta t_e$ and  received at $r=0$ at $ t_o+\delta t_o$. In 
the same way, one  obtains 
\begin{equation} 
\int_{t_e +\delta t_e}^{t_o + \delta t_o} \frac{\d t}{a(t)} 
=- \int_{r_e}^{0}\d r \,.  \label{radialintegral2} 
\end{equation} 
Since the right hand sides  of eqs.~(\ref{radialintegral1}) and 
(\ref{radialintegral2})  are  equal, so are their left hand 
sides, 
\begin{equation} 
\int_{t_e +\delta t_e}^{t_o + \delta t_o} 
\frac{\d t}{a(t)} = \int_{t_e }^{t_o} \frac{\d t}{a(t)} 
\,, 
\end{equation} 
and one can then write 
\begin{equation} 
\left[ \int_{t_e +\delta t_e}^{t_o }
+ \int_{t_o}^{t_o +\delta t_o} - \left( \int_{t_e}^{t_e 
+\delta t_e} +\int_{t_e   +\delta t_e}^{t_o} \right) \right] 
\frac{\d t}{a(t)} =0 
\end{equation} 
and 
\begin{equation} 
\int_{t_o}^{t_o+\delta t_o} \frac{\d t}{a(t)} = 
\int_{t_e}^{t_e+\delta t_e} \frac{\d t}{a(t)} \,. 
\end{equation} 
Assume now that $\delta t_e$ and $\delta t_o$ are very small,  so 
that $a(t)$ does not change appreciably from its value $ 
a(t_e)$ [respectively,   $a(t_o)$] in the time interval $\left( 
t_e ,t_e+\delta t_e \right)$  [respectively, $\left( t_o 
  ,t_o+\delta t_o \right)$]; then 
\begin{equation} 
\frac{\delta t_o}{a(t_o)}= \frac{\delta t_e}{a(t_e)} 
\end{equation} 
and, assuming now $ \nu_e=1/\delta t_e $ to be  the frequency of 
the signal at emission and $ \nu_o=1/\delta t_o $ the received 
frequency, both measured with respect to the fundamental frequency 
unit $1/\tilde{t}_{u}$, it is 
\begin{equation} 
\frac{1}{a(t_o) \nu_o}= \frac{1}{a(t_e) \nu_e} \,. 
\end{equation} 
The redshift $z$ is then given by 
\begin{equation} 
z+1 \equiv  \frac{\lambda_o}{\lambda_e} = 
\frac{\nu_e}{\nu_o}=\frac{ a(t_o )}{a(t_e )} \equiv 
\frac{a_o}{a_e} 
\end{equation} 
(where $\lambda_e$ and $ \lambda_o$ are the wavelengths at 
emission and observation, respectively). Then there is redshift 
also in   flat ``veiled frame'' space and its derivation 
parallels completely  the standard derivation of redshift in 
FLRW space  ({\em e.g.}, \cite{dInverno}). The result agrees 
with Ref.~\cite{Deruelle:2010ht}, in which the cosmological 
redshift in the veiled frame is derived in a different way by 
considering an hydrogen atom and taking into account carefully 
the local change in the electron mass deriving from the 
non-trivial coupling of matter to the conformal factor $W$ in the 
veiled frame. This coupling can also be interpreted as a  
variation of units with the spacetime point \cite{Dicke} and it 
is the source of redshift. The distance-redshift relation in the 
veiled frame is also derived in Ref.~\cite{Deruelle:2010ht}, and 
it coincides, of course, with the one derived in FLRW space with 
constant units.

In the flat space of veiled FLRW cosmology the radial null 
vectors have components (we can use the coordinates $\eta$ and $r$ from 
(\ref{manifestflrw}) 
if $k=0$) \beq l^{\mu} = \left( 1, 1, 0 ,0 \right) \,, 
\;\;\;\;\;\;\;\; n^{\mu} = \left( 1, -1, 0 ,0 \right) \,. \eeq 
The expansions of these two null vectors in the flat space are 
\beq \theta_{l} = \frac{2}{r} \,, \hspace{4cm} \theta_{n} = 
-\frac{2}{r} \,. \eeq We have $\theta_{l}\theta_{n}<0$ for all 
finite $r$, so there are no spherical trapped surfaces in flat 
space (in fact there are no trapped surfaces entirely contained 
in flat space at all). But if we instead look at the change of 
the horizon-entropy in the null directions we find \beq 
l^{a}\nabla_{a}\left(a^{2}A\right) = \partial_{\eta}(a^{2}A) + 
\partial_{r}(a^{2}A) = 8\pi ar\left[ \left(\partial_{\eta}a 
\right)r + a\right] \,, \eeq

\beq n^{a}\nabla_{a}\left(a^{2}A\right) = \partial_{\eta}(a^{2}A) 
- \partial_{r}(a^{2}A) = 8\pi ar\left[ \left( \partial_{\eta}a 
  \right)r - a\right] \,. \eeq We see that we have a conformal 
  horizon at $r=a/(\partial_{\eta}a)$. We can convert the 
  coordinates from $\eta$ to $t$ by writing $ \partial_{\eta}a = 
  a\partial_{t}a$. Thus, there is a horizon at $ 
  r=1/(\partial_{t}a)$ just as in the spatially flat FLRW case. We can also 
compute 
\begin{eqnarray} 
l^{a}\nabla_{a}\left[ n^{b}\nabla_{b}(a^{2}A)\right] & = & 
8\pi  ar^2\partial_{\eta}\partial_{\eta}a = 8\pi a^{3} 
  r^{2}\left(\ddot{a}+\frac{\dot{a}^{2}}{a}\right) \nonumber\\
&&\nonumber\\ 
&=& 8\pi a^{4}r^{2}H^{2}(1-q) \,. 
\end{eqnarray} 
We obtain exactly the same kind of horizon as above (a Past  
Inner Trapping Horizon in the case of de Sitter). The signature 
of the horizon is the same (as expected) since the normal is 
\beq 
N_{\mu} = \left(\frac{a\ddot{a}}{\dot{a}^{2}}, 1, 0, 
  0\right) \,, 
\eeq 
whose norm is given by
\beq 
N^{a}N_{a} = 1-\left(\frac{a\ddot{a}}{\dot{a}^{2}}\right)^{2} = 
1-q^{2} \,. 
\eeq
It may come to a surprise, especially for an 
  astronomer, that the Hubble parameter (which is also three 
  times the expansion of the timelike worldlines of comoving 
  observers and a scalar quantity) is not a good observable when 
  studying cosmological horizons and their location (see, {\em 
  e.g.}, \cite{Capozzielloetal10}). However, from the discussion 
  above, it is clear that $H$ is not a good observable when 
  conformal transformations are used in generalized (and even in 
  Einstein) gravity. $H$ is changed by conformal transformations 
  and so is the location of the cosmological horizon, and a more 
  general quantity is needed.

%************************************************************* 
\section{Conclusion} 
%************************************************************* 
\label{sec:conclusion}
If entropy is a useful quantity in time-dependent 
situations, and possibly also in non-equilibrium thermodynamics, its applicability may extend beyond event 
horizons of static or stationary black holes. Dynamical 
situations are the rule rather than the exception and, in certain 
theories, stationary situations may not even exist. For example, 
in the class of $f(R)$ theories designed to explain the present 
acceleration of the universe without resorting to dark energy,
 Minkowski space is not a global solution and one cannot 
contemplate asymptotically flat black holes in these theories. 
When the relevant 
field equations are written in a form that mimics the Einstein 
equations, a cosmological effective fluid composed of geometric 
terms is present on the right hand side of these equations, and 
causes the universe to accelerate its expansion, so that the role 
of Minkowski space as a global solution giving a static 
background is played instead by the de Sitter or other FLRW 
solutions. In this case, black holes are embedded in dynamical 
(cosmological) backgrounds and one does not have the luxury of 
considering static horizons in a static background. However, the horizon-entropy formula (\ref{waldentropy}) originally developed for perturbations of stationary systems, gives rise to rather generic horizon-entropy increase laws for both causal and quasi-local horizons in general dynamical spacetimes.

Here we have seen how a modification of the trapping horizon 
conditions can give quasi-local horizons for which a horizon-entropy increase 
law can be proven in models that are related via field 
redefinitions to Brans-Dicke theory. The location of 
these surfaces is invariant under a conformal transformation of 
the metric, which is not true of ordinary trapping horizons. It is likely that these results will hold for all 
theories that are conformally related to Einstein gravity and for which the horizon-entropy transforms in the 
same way as the area. These 
conditions can be applied to a variety of situations including 
finding cosmological horizons in ``veiled'' Minkowski space 
with varying units.

In a given spacetime there are many surfaces for which one can 
define an entropy increase law. We have examined here three 
different cases, null causal horizons which include global event 
horizons, locally defined geometric horizons including trapping 
horizons, and the new proposal based on gravitational horizon-entropy. We have derived quasi-local conditions on the rate of increase of the horizon-entropy and shown that this is non-negative for both causal horizons and the new quasi-local horizons. Although the governing equation for these two cases 
is very similar, compare
 eq.~(\ref{seconddifflns}) and eq.~(\ref{entvaryconf}), there are 
 some differences. The main difference is that in the case of 
 causal horizons it governs the behaviour of the second 
 derivative of the horizon-entropy and for the quasi-local horizons it 
 governs directly the first derivative of the entropy. The 
 horizon-entropy of both types of horizons can shrink if sufficient 
 negative energy is provided. Both types of horizons can settle 
 down to exact Killing horizons, but only the quasi-local 
 horizons can start from exact Killing horizons. While $\omega_{l}=0$ 
 follows trivially for causal horizons, for
quasi-local horizons it requires the additional assumption that 
the null normal $l^a$ is derivable from a double-null foliation.

We have proven that horizon-entropy does not decrease by 
requiring that all the individual terms in (\ref{seconddifflns}) and (\ref{entvaryconf}) 
are negative. In particular this requires the null energy 
condition $T_{ab}l^{a}l^{b} \geq 0$ be satisfied for matter 
fields, rather than the null curvature condition 
$R_{ab}l^{a}l^{b} \geq 0$. In fact, all that is actually 
required is that the overall sum of the terms in (\ref{seconddifflns}) or
(\ref{entvaryconf}) be negative. It is possible that in certain 
specific scenarios some of these terms, in particular 
$-T_{ab}l^{a}l^{b}$,  are positive, but that overall the 
horizon-entropy still increases.

The quasi-local surfaces used to derive the horizon-entropy 
increase law are in general not apparent horizons or trapping 
horizons. Outside of the Einstein frame  they are not foliated by 
marginally outer trapped 
surfaces except in the case where they describe Killing or 
isolated horizons and {in general} they do not 
satisfy $\theta_{l} = 0$. They will though be spacelike surfaces 
if the horizon-entropy is increasing and null surfaces if it is 
constant. In a spacetime that satisfies the null energy condition 
they will be located behind the event horizon and so will, in 
cases like the Brans-Dicke collapse considered in 
\cite{Scheel:1994yn}, lie inside the apparent horizon.

It was mentioned in \cite{Ford:2000xg} that apparent horizons 
will not satisfy a horizon-entropy increase relation and that the 
acausal behaviour of event horizons is needed to save this law. 
The surfaces given by (\ref{modtraphorconds}) are quasi-locally 
defined and satisfy a local horizon-entropy increase law of the 
form used in \cite{Ford:2000xg}. In \cite{Fiola} the validity of 
the Generalised Second Law (GSL) was examined for apparent 
horizons in a string frame two-dimensional model. This work 
explicitly included the contribution of both the horizon-entropy 
and the entropy of fields outside the horizon and concluded that for 
coherent quantum states the GSL was valid but possibly violated 
for non-coherent quantum states. It is not known whether the 
surfaces satisfying (\ref{modtraphorconds})  will satisfy the 
GSL. Since they coincide with trapping horizons in the Einstein 
frame, if it can be shown that the GSL is violated for trapping 
horizons in the Einstein frame then the same will be true for 
these surfaces.

We have used a dynamical definition of entropy as proposed in 
\cite{Wald:1993nt}. Strictly speaking this definition is derived 
only for stationary situations and it is known that its 
application to non-stationary situations contains several 
ambiguities \cite{Jacobson:1993vj, Iyer:1994ys}. These 
ambiguities are not essential for our derivation, in fact all we 
require is a definition of horizon-entropy whose value is 
invariant under a conformal transformation.  Even the association 
of $s_{ab}$ with entropy is not essential,  only that it 
transforms invariantly under a conformal transformation. Throughout this work we have suppressed the factor of $4$ in the area-entropy relation, and our results are independent of the precise numerical value of this factor. 

Under a conformal transformation of the metric, the location of 
the surfaces studied here remains the same. This is not true of 
trapping horizons. That the surfaces are invariant under a 
conformal transformation, is in a certain sense trivial, because 
the horizon-entropy definition used is always equal to the area in the 
Einstein frame and so the definitions always pick out the 
``Einstein frame trapping horizon''. Put simply, we have \bea 
g_{ab} \rightarrow \tilde{g}_{ab} & = & W g_{ab} \,,\nonumber \\ 
A \rightarrow \tilde{A} & = & WA \,, \nonumber \\ S \rightarrow 
\tilde{S} & = & \frac{\tilde{A}}{4W} = \frac{A}{4} = S \,. \eea 
But the conditions (\ref{modtraphorconds}) do not make explicit reference to 
the Einstein frame and thus can be applied simply in non-Einstein 
frames without the need to transform the metric. While a 
conformal transformation will always put the theories considered here 
into the Einstein frame form of Einstein gravity plus matter, and one could proceed with traditional trapping horizons, one 
must accept that in many alternative theories of gravity this 
Einstein frame will not be the standard frame with constant 
units.

We have argued, along with many other authors, that a conformal 
transformation of the metric should not change the operationally 
defined physical features of the spacetime, provided that one 
redefines standards of length, time and mass in a 
position-dependent way. This is most easily demonstrated in the case of ``veiled general relativity'' where metric solutions of ordinary Einstein gravity are subjected to a conformal transformation. In the simple case of the Schwarzschild spacetime the usual conditions for a trapping horizon do not always pick out the $r=2M$ surface. The modified conditions proposed here do. Thus, the surfaces defined here allow a more operationally physical interpretation than trapping horizons.

\ack A.B.N. is very grateful for generous support from the 
Alexander von Humboldt Foundation and hospitality at the Max 
Planck Institute for Gravitational Physics in Potsdam-Golm and 
Bishop's University. V.F. acknowledges financial support from 
Bishop's University and the Natural Sciences and Engineering 
Research Council of Canada.  
\end{document}